\documentclass[11pt]{article}

\usepackage{pifont}
\usepackage{graphicx}
\newcommand{\be}{\begin{equation}}
\newcommand{\ee}{\end{equation}}
\newcommand{\bea}{\begin{eqnarray}}
\newcommand{\eea}{\end{eqnarray}}
\newcommand{\twospinor}[2]{  \left( \begin{array}{ccc}{#1} \\ [6pt] {#2} \end{array} \right) }

\begin{document}

%\begin{frontmatter}

%\title{Elsevier \LaTeX\ template\tnoteref{mytitlenote}}
%

\title{Electron internal energy and internal motion (Zitterbewegung) as consequence of local U(1) gauge invariance in two-spinor language}
\maketitle


\author{J. Buitrago}

%\address
{University of La Laguna, Faculty of Physics}

{38205, La Laguna, Tenerife, Spain}

%\begin{keyword}
%
%
%Weyl 2-spinor formalism; Gauge Invariance ; Dirac Equation; Lorentz Force;
%\end{keyword}



%\ead
{jgb@iac.es}

%
%\end{frontmater}


%\begin{abstract}
%This template helps you to create a properly formatted \LaTeX\ manuscript.
%\end{abstract}
%
%
%********************************************************************
%

%\email{jgb@ll.iac.es}
%\date{\today}

%\pacs{03.30.+p, 04.20.
%\maketitle
%

%

% First modification to the English 26.04.2018  Write ``associated with'' instead of ``associated to''.  
\begin{abstract}
 Starting with the results obtained in a previous paper in which classical local U(1) gauge invariance in terms of the electromagnetic field strenghts instead of the usual formulation mediated by the four potential was introduced  it is shown that using the gauge freedom associated with the third component  of the magnetic field, previously obtained spinor equations of motion describe the internal dynamics of a free 1/2 spin particle suggesting a kinematic origin of its rest mass and helicity. The controversial Zitterbewegung (trembling motion) appears in a natural way as internal motion with the velocity of light. Such an interpretation is in contrast with the usual quantum mechanical explanation of transitions between positive and negative energy states.

\end{abstract} 

%
%\maketitle

%
%\end{frontmatter}

%

\section{Introduction}

The main purpose of this article is to present the  Zitterbewegung (trembling motion) phenomena from a different perspective and other outlook than the usual quantum mechanical interpretation based in the Dirac equation. As far as I know, the only study leading to the same results and similar interpretation dates back to 1990 (see \cite{hestenes}). Curiously enough both approaches being conceptually and technically quite different converge to the same results.
From the technical and conceptual standpoint, the present work  is based on a previous paper \cite{bui}  presenting a  {\it classical} $U(1)$ local gauge invariance formulation via a lagrangian with an interaction in terms of the electric and magnetic field strengths. This approach is certainly different to the usual lagrangian leading to the Dirac Equation interacting with an external electromagnetic field described by the four potential $A^{\mu}$. 
% two more modifications to the language 26.04.2018. Use the word ``duplication'' instead ``duplicity'' because ``duplicity'' means deceptiveness. Also corrected spelling of ``continuous''
As the present study is strongly based on the mentioned article, I have tried to find a compromise between continuous referencing to [1] and a certain self consistency while avoiding too much duplication.

%\section{U(1) Gauge invariance in Weyl two-spinor form} 

The starting point of \cite{bui} (see also \cite{buitrago}) are the following linear first order differential spinor equations (the derivative is taken respect to the proper time $\tau$):
\begin{equation}\label{eqofmot}
\begin{array}{c}
\frac{d\eta^A}{d\tau}=\frac{e}{m}\phi^{AB}\eta_{B} \\
\frac{d\pi_A}{d\tau}=-\frac{e}{m}\phi_{AB}\pi^B \\
\end{array}
\end{equation} \\
%
% Corrected spelling of ``typically'' 26.04.2018  
(A short derivation of these equations is included at the beginning of next section) These coupled spinor equations (natural units $\hbar=c=1$ will be used) are equivalent to the Lorentz Force describing the motion of a 1/2 spin particle  of mass $m$ and charge $e$ (typically an electron) under an electromagnetic field described by the symmetric second-rank spinor $\phi_{AB}$,  explicitely given by
\begin{equation} \label{phi}
\phi_{AB}= \frac{1}{2}
\left( \begin{array}{cc}
\left[ E_{1}-iB_{2}\right] -i\left[E_{2}-iB_{2}
\right] & -E_{3}+iB_{3} \\
-E_{3}+iB_{3} & \left [-E_{1}+iB_{2}\right]-i
\left[E_{2}-iB_{2}\right] \\
\end{array}\right) .
\end{equation}
In turn, $\phi_{{AB}}$ and its complex conjugate $\bar\phi_{{A'}{B'}}$ form the antisymmetric four-rank electromagnetic field spinor in its standard form that can be found in \cite{penrose} and any other book dealing with this subject.
\begin{equation} \label{emfield}
F_{ABA'B'}=\epsilon_{AB}\bar\phi_{A'B'}+
\epsilon_{A'B'}\phi_{AB},
\end{equation} \\
where $\epsilon_{AB} = \epsilon^{AB}$ is the spinor metric
\begin{equation}\label{metric1}
\epsilon_{AB}=\left(\begin{array}{cc} 0 & 1 \\ -1& 0 \end{array}\right).
\end{equation} \\
%
%
% 19.06.2018 Replace `Capital index (singular) ' with `Capital indices (plural)' 
(Capital indices are lowered (raised) by means of the metric spinor $\epsilon_{AB}$ already defined above. See Appendix 1)

The solution of equations (\ref{eqofmot}) determine the 
four momentum of the particle given by the hermitian spinor defined  as superposition of the two null directions $\pi^A\bar\pi^{A'}$ and $\eta^A\bar\eta^{A'}$ as \cite{bette}
\begin{equation} \label{momentum}
p^{AA'}=\frac{1}{\sqrt{2}}\left[\pi^{A}\bar\pi^{A'}+\eta^{A}\bar\eta^{A'}\right].
\end{equation} \\
Since $p^{AA'}$ is to represent the four-momentum of a massive particle, must be time-like and certainly fulfill the condition:
\begin{equation} \label{condition}
p^{AA'}p_{AA'}=m^2.
\end{equation} \\
% 19.06.2018  I recommend inserting the diagram here. 
%\begin{figure}[h]
%\begin{center}
%\includegraphics [width=1.0 \textwidth]{figure-1.pdf}
%\caption{If $\eta^A$ and $\pi^A$ are two non proportional spinors, the four momentum of a fermionic particle of mass $m$ can be defined as the superposition of the two null directions, $\eta^A \eta^{A'}$ and $\pi^A \pi^{A'}$.(A. Bette, J.Math.Phys 34, 1993)  } 
%\end{center}
%\end{figure}

On the other hand, following the standard representation, the different components of $p^{AA'}$ are labeled according to
\begin{equation} \label{strep}
p^{AA'}=\left( \begin{array}{cc}
p^{00'} & p^{01'}\\
p^{10'} & p^{11'} \\
\end{array}\right)
= \frac{1}{\sqrt 2}\left( \begin{array}{cc}
p^0+p^3 & p^1+ip^2 \\
p^1-ip^2 & p^0-p^3 \\
\end{array}\right).
\end{equation}
This last expression must be used when solving an specific case, via the spinor equations, to identify the components of $p^{AA'}$ in the solution.

\section{$U(1)$ Local gauge transformation of field strength quantities and Zitterbewegung}
%  19.06.2018  Replace `at' the introduction with `in' the introduction 
As I have mentioned in the introduction, the particle field interaction, will not involve the four potential $A^{\mu}$ as in the traditional approach. Instead, the coupling will be to the electric and magnetic field strengths. To this end and to emphasize the geometrical origin of this formulation (see also \cite{buitrago}) consider an infinitesimal transformation of $\eta^{A}$ mediated by an element of the group SL(2,C):
\begin{equation}
\delta\eta^{A}=exp\left [\frac{1}{2}(\vec\delta\omega\vec\sigma+i\vec\delta\theta\vec\sigma)\right]\eta^{A},
\end{equation} \\
being $\vec\delta\omega$ and $\vec\delta\theta$ the infinitesimal parameters of boosts and rotations respectively and $\vec\sigma$ the usual Pauli matrices.
Since the transformation is infinitesimal, we can write
\begin{equation}
\delta\eta^{A}=\left[ I+\frac{1}{2}(\vec\delta\omega\vec\sigma+i\vec\delta\theta\vec\sigma)\right]\eta^{A}
\end{equation}
% 19.06.2018  Surely you want to cite the 1995 and 2007 papers rather than the 2016 paper here (?)  
According to the interpretation given in \cite{buitrago} associating the fields $\vec E$ and $\vec B$ with infinitesimal boosts and rotations, and via a classical detour, we would say that any change in the dynamic state of the particle should be proportional to the force field acting on the particle and the lapse of proper time. Accordingly 
\begin{equation}
\vec\delta\omega=K \vec\epsilon(x)\delta\tau
\end{equation}
\begin{equation}
\vec\delta\theta=K \vec\beta(x)\delta\tau
\end{equation}
By substitution of $\epsilon$ and $\beta$ by $\vec E$ and $\vec B$ and expanding the term associated with the Pauli matrices we get a second rank spinor (coincident with the second rank electromagnetic field spinor $\phi^A_{\ B}$ ):
\begin{equation}
\phi^A_{\ B}= \frac{1}{2}\left[\left(\begin{array}{cc} E_3& E_1+iE_2 \\ E_1-iE_2& -E_3\end{array}\right)+i\left(\begin{array}{cc} B_3& B_1+iB_2 \\ B_1-iB_2& -B_3\end{array}\right)\right].
\end{equation}
It is only necessary to lower the first upper index, following the rules given in the first appendix, to find the symmetric form (\ref{phi}) of the field spinor $\phi_{AB}$. On the other hand, from equation (9) and subsequents equations, making $K=e/m$, it is immediate to obtain
\begin{equation}
\frac{d}{d\tau}\eta^A=\frac{e}{m}	\phi^A_{\ B}\eta^B.
\end{equation}

As was done in [1] and repeated here for self-consistency, the lagrangian density, for a free particle, along the {\it classical} path of the particle (with dimension energy per unit length) is 
% 23.08.18 Use ``separate'' rather than ``separated'' in this context, i.e in this footnote.
% Also replace ``have'' by ``has''  in  
\footnote{In the definition of the four-momentum $p^{AA'}$, the spinors $\pi^A$ and $\eta^A$ enter on equal footing. It is then clear that the lagrangian could also be defined swapping both spinors.}
%or even as ${\cal L}= \pi_A \dot\eta^A+\eta_A \dot\pi^A$ which would involve two separate Euler-Lagrange equations to obtain both equations of motion. The motivation for this choice has been only to keep it short (see Appendix 2)}.
%
%\footnote{There is a certain analogy with the lagrangian density leading to the Dirac Equation for a free particle: ${\cal L}=i\bar\psi\gamma^\mu\partial_\mu\psi- m\bar\psi \psi$. Note, however, that, apart from dealing here with a classical description, the more significative difference is the absence of a mass term which is due to the non hamiltonian approach. The mass is introduced via the U(1) local gauge invariance principle in the $e/m$ ratio term characteristic of electromagnetic forces. Note also that Dirac spinors have four components thus differing from Weyl two-spinors. In Weyl version of the Dirac equation the gamma matrices are absent \cite{penrose}}

\begin{equation} \label{lagrf}
{\cal L}=\pi_A \dot{\eta}^{A},
\end{equation} \\
% 23.08.2018 Have removed the ``s'' at the end of equations as this refers to a single equation. 
(dot denoting derivative respect to proper time $\tau$) together with the Euler Lagrange equation
%
% 19.06.2018 At this point I think that more explanation for thr form of \ref{lagrf} is needed. 
\begin{equation} \label{1.3}
\frac{d}{d\tau}\frac{\partial \cal L}{\partial \dot{\eta}^{A}}-\frac{\partial \cal L}{\partial \eta^{A}}=0.
\end{equation} \\
% 23.08.2018 insert ``a'' and remove ``s'' as again this is a single equation (isn't it?) 
With a similar equation for the spinor $\pi^A$. The former equation leads to $\dot{\pi}_{A}=0 \Rightarrow \pi_{A}=const.$\\
We consider now the consequences of imposing invariance under local ({\it along the classical path parametrized by $\tau$}) phase transformations
\begin{equation} \label{1.6}
\begin{array}{c}
\eta^{A} \rightarrow e^{i\alpha(\tau)}\eta^{A} \\
\pi^{A} \rightarrow e^{i\xi(\tau)}\pi^{A}. \\
\end{array}
\end{equation} \\
%

%.
The phase parameters $\alpha(\tau)$ and $\xi(\tau)$ cannot be independent as the spinors $\eta^A$ and $\pi^A$ are also not independent since from (\ref{momentum}), they are related by the condition \\
\begin{equation} \label{1.7}
p^{AA'}p_{AA'}=\vert\pi^{A}\eta_{A}\vert^{2}=m^{2}.
\end{equation}
In consequence $\eta^A\pi_A=const.$, leading to the constraint $\xi(\tau)=-\alpha(\tau)$.\\

As the classical trajectory should not be affected by any phase transformation, it is apparent that local gauge transformations leaves invariant the four-momentum of the particle:
\begin{displaymath}
p^{AA'}=\frac{1}{\sqrt{2}}\left[\pi^{A}\bar{\pi}^{A'}+\eta^{A}\bar{\eta}^{A'}\right].
\end{displaymath} \\
%   23.08.2018 Have added ``s'' to ``transform'' 
However, the free lagrangian (\ref{lagrf}) transforms to \\
\begin{equation} \label{1.10}
{\cal L} \rightarrow i\dot{\alpha}\eta^{A}\pi_{A}+
\dot{\eta}^{A}\pi_{A}=i\dot{\alpha}\epsilon_{AB}\eta^B\pi^A+\dot{\eta}^A\pi_A.
\end{equation} \\
To find a gauge invariant lagrangian we have to add a term
\begin{equation} \label{1.12}
-\frac{e}{m}\phi_{AB}\eta^{B}\pi^{A},
\end{equation} \\
and impose the condition for the new field $\phi_{AB}$ of transforming, under local phase
transformations, as 
%corrected spelling of name of author of book to Stewart J 26.04.2018  
\footnote{From a pure mathematical point of view, the validity of transformation (20) is consequence of the following theorem applied to valence-2 spinors (see Stewart J. {\it Advanced General Relativity}. 1991 Cambridge Univ. Press. Page 69): ``Any spinor $\tau_{A...F}$ is the sum of the totally symmetric spinor $\tau_{(A...F)}$ and (outer) products of $\epsilon 's$ with totally symmetric spinors of lower valence}\\

\begin{equation} \label{1.13}
\phi_{AB} \rightarrow \phi_{AB}+i\frac{m}{e}\dot{\alpha}\epsilon_{AB},
\end{equation} \\
then, the new lagrangian \\
\begin{equation} \label{1.14}
{\cal L}=\dot{\eta}^{A}\pi_{A} -
\frac{e}{m}\phi_{AB}\eta^{B}\pi^{A},
\end{equation} \\
is invariant under $U(1)$ local-phase transformations. The transformation 
that holds
for the conjugate second-rank spinor   $\bar{\phi}_{A'B'}$,  is
given by \\
\begin{equation} \label{1.15}
\bar{\phi}_{A'B'} \rightarrow \bar{\phi}_{A'B'}-i\frac{m}{e} \dot\alpha\epsilon_{A'B'}.
\end{equation} \\
These kind of transformations leave however
invariant the associated
four-rank spinor of the Maxwell field strength \cite{penrose} \\
\begin{displaymath}
F_{ABA'B'}=\epsilon_{AB}\bar{\phi}_{A'B'}+\epsilon_{A'B'}\phi_{AB}.
\end{displaymath} \\
%
%In general, $F_{ABA^'B^'}$ is invariant under any %transformations of the form:\\
%
%See comments in pag 24 of spinorials equations\\
From the Euler Lagrange equations applied to the lagrangian given by (\ref{1.14}) it is immediate to obtain
\begin{equation} 
\dot\pi_A=-\frac{e}{m}\phi_{AB}\pi^B.
\end{equation}
This equation and those of   (\ref{eqofmot}) are gauge invariant.
% 23.08.2018 have replaced modify with modifies 
If the invariance of the four-rank spinor $F_{ABA'B'}$ follows from the transformation rules of the field spinors $\phi_{AB}$ and $\bar\phi_{A'B'}$ a simple look at (\ref{phi}) and (\ref{metric1}) reveal that only the components $\phi_{01}$ and  $\phi_{10}$ (together with their complex conjugates) are affected. Furthermore since the transformation is purely imaginary, there is only one field quantity (i.e. $ B_3$) affected by the transformations. Consequently, the spinor equations are modified in much the same way as a gauge change in the electromagnetic potential $A^\mu$ modifies the equations and their solutions (for example: the Coulomb gauge in QED). 
It is clear that this peculiarity of the third component of the magnetic field deserves further attention.
In what follows, we shall examine the consequences of this local gauge invariance for $B_3$. For simplicity the case of $B_3=const.$ will be studied. 
Given the equation

\begin{equation} 
\dot\eta_A=-\frac{e}{m}\phi_{AB}\eta^B,
\end{equation}
and applying the gauge transformations:
\begin{equation} 
\phi_{01} \rightarrow \tilde\phi_{01}=\phi_{01}+i\frac{m}{e}\dot\alpha\epsilon_{01}
\end{equation}
\begin{equation}
\phi_{10} \rightarrow \tilde\phi_{10}=\phi_{10}+i\frac{m}{e}\dot\alpha\epsilon_{10}
\end{equation}
The transformed equations for the physical components (upper indices) are
\begin{equation}
\dot\eta^0 = \frac{e}{m}i \left(\frac{-B_3}{2} + \frac{m}{e}\dot\alpha \right)\eta^0
\end{equation}
\begin{equation}
\dot\eta^1 = \frac{e}{m}i \left(\frac{B_3}{2} + \frac{m}{e}\dot\alpha \right)\eta^1.
\end{equation}
As in the last equations we cannot gauge away simultaneously both effective fields, let us choose
\begin{equation}
-\frac{B_3}{2} +\frac{m}{e}\dot\alpha =0,
\end{equation}
then
$$\dot\eta^0 = 0 \Longrightarrow \eta^0 = const.= \sqrt {\hat m}$$
%
% Replaced ''need`` with ''needs to``   26.04.2018 
(the integration constant $\sqrt{\hat m}$ needs to have the dimension of an energy squared)
Under condition (29), (28) reduce to
\begin{equation}
	\dot\eta^1 = \frac{e}{m}i \left(\frac{2m}{e}\dot\alpha \right)\eta^1 = i2\dot\alpha\eta^1
\end{equation}
As  both charge and mass of the particle as well as magnetic field have disappeared, the last relation should be valid for any 1/2 spin particle. Doing the integration:
\begin{equation}
\eta^1= \sqrt{\hat m} e^{i2\dot\alpha\tau}
\end{equation}
The spinor $\eta^A$ is 
\begin{equation}
\eta^A = \sqrt{\hat m} \left(\begin{array}{c}
1 \\
e^{i2\dot\alpha\tau} \\
\end{array} \right).
\end{equation}
As for the other spinor and proceeding in the same way:
\begin{equation}
\pi^A = \sqrt{\hat m} \left(\begin{array}{c}
1 \\
e^{i2\dot\alpha\tau} \\
\end{array} \right).
\end{equation}
%
%The four momentum is defined (see [1]) as
Denoting $\dot\alpha=\omega$,
from the four-momentum definition (\ref {momentum}) and (\ref{strep})
\begin{equation} \label{strep1}
p^{AA'}=\frac{1}{\sqrt 2}\left( \begin{array}{cc}
2\hat m & 2 \hat m e^{i2\omega\tau} \\
2\hat m e^{-i2\omega\tau} & 2\hat m \\
\end{array}\right).
\end{equation}
Remembering that $Det [p^{AA'}]$ equals 1/2 the Lorentz norm, we have a null path in momentum spacetime.
After some lengthy calculations to solve for the components:
$$
p^0 = E =2\hat m
$$
$$
p^3 = 0
$$
$$
p^1 = 2\hat m \cos 2\omega\tau
$$
$$
p^2 = - 2\hat m \sin 2\omega\tau .
$$
$2mc^2/ \hbar$ (in conventional units) is the so called Zitterbewebung (trembling motion), found in the Dirac Theory and subject of many controversial interpretations. From the last and following relations, we shall find a classical interpretation as an internal circular motion taking place inside of the particle (electron). It makes sense now to identify $\hat m$
as the electron mass.
Since $p^i = 2mu^i$, (i = 1,2)
\begin{equation}
	(p^1)^2 + (p^2)^2 = 4m^2 \Longrightarrow (u^1)^2+(u^2)^2 = 1.
\end{equation}
(Circular motion in the x-y plane with the velocity of light).

Let us now take the other alternative (20), namely
\begin{equation}
\frac {B_3}{2} + \frac {m}{e}\dot\alpha = 0
\end{equation}
The solutions are now
\begin{equation}
\pi^A = \sqrt{m} \left(\begin{array}{c}
e^{i2\dot\alpha\tau} \\
1
\end{array} \right)
\end{equation}
\begin{equation}
\eta^A = \sqrt{m} \left(\begin{array}{c}
e^{i2\dot\alpha\tau} \\
1
\end{array} \right).
\end{equation} 
Performing the same calculations as in the other case
$$
p^1 = 2m \cos 2\omega\tau
$$
$$
p^2 = 2m \sin 2\omega\tau
$$

The spatial trajectories in the x-y plane can be immediately obtained
:

$$
x = \frac {1}{2\omega}\sin 2\omega \tau $$
$$
y = -\frac {1}{2\omega}\cos 2\omega \tau 
$$
and
$$
x = \frac {1}{2\omega}\sin 2\omega \tau $$
$$
y = \frac {1}{2\omega}\cos 2\omega \tau 
$$

Being clock and counter clock-wise respectively.

The previous results concerning the momentum and position of the electron behavior are in agreement with those obtained in \cite{hestenes} (see equation 64 in the cited work).
\section{Discussion}
% 19.06.2018  Replace `could be' with `may be' and replace correct spelling of `opposite' 
% 19.06.2018  Replace `as early of 1930' with `as early as 1930' 
% Replaced a double quote symbol '' (above the 2 on my keyboard) with one single `s on the far left of my keyboard and a single ' below the @ symbol on my keyboard around ``e'' and ``Zitterbewegung''. 
A fundamental question that, perhaps, we have some reasons to ask now is what is the origin of the electric charge. Previously the charge ``e'' disappeared in (23) and did not appear again. However, the results seems independent of the charge. Since the neutrinos have a certain mass, and possibly a magnetic moment, it may be possible to explain the opposite helicities of neutrinos within the three families. As early as 1930, Schrodinger \cite{sch}  made the first analysis of what was to be called afterwards ``Zitterbewegung'' (Trembling Motion). The frequency $2\omega$ also appears in the solutions of the Dirac Equation for the propagation of a free packet. He interpreted $\omega_0 = 2mc^2/ \hbar$ as a fluctuation in the position of the electron with radius 
$$
\Delta r =\frac{c}{\omega_0} = \frac{\hbar}{2mc}.
$$
Assuming a velocity of the electron equal to the velocity of light about some mean position inducing an spin angular momentum
$$
\Delta r . mc = \frac {\hbar}{2}.
$$
%
% Use either ``relatively modern'' or ``more modern'' 26.04.2018    
A relatively modern interpretation \cite{bjork} is that there are unavoidably cross terms between the positive and negative energy solutions which oscillate rapidly in time with frequencies
$$
\frac{2p_0 c}{\hbar} \geq \frac {2mc^2}{\hbar}
$$

%
%  Replace ``There is nothing special in..'' with ``There is nothing special about...'' 26.04.2018
% Also replace ``have'' with ``has'' in the z-axis have, in practice a (not ``an'') special relevance... 
Perhaps the reader will ask why $B_3$? The answer has to do with the choice of the third Pauli matrix $\sigma_3$. There is nothing special about the $z$-axis, but once we choose this axis for  $\sigma_3$, the $z$-axis has, in practice, a special relevance. (More technically, magnetic fields are related to rotations described by the SO(3) group having SU(2) as covering group (see \cite{bui0} , [1] and \cite{physreva} ).

Finally, and as already mentioned, the results in this work seems to be in agreement with those obtained in [2].
\
\
%

% {\noindent\bf Appendix 1}

% 19.06.2018  I suggest an enlargement of the appendix to aid clarity. My version of the appendix in which I have incorporated material from the 2017 conference talk is given in `comments_on_revised_version.pdf' and of course `comments_on_revised_version.tex'. TO SAVE TIME, I have commented out the version here and inserted the enlarged version rather inserting modifications into the existing text individually.

% So immediately below is my enlarged version 
\bigskip

{\noindent\bf Appendix 1: A Short Introduction to Spinor Calculus}
%

% 19.06.18  Have replaced double quotes with two single quotes around ``Spinors and space-time'' and also included citation here. Have also corrected spelling of `accessible'. 
With the aim of making this article accessible to potential readers not familiar with two-spinor formalism, I consider that
a very simple and basic introduction to spinor calculus would be particularly helpful for the obvious reason that the formalism originally developed by Penrose and Rindler in their books ``Spinors and space-time'', \cite{penrose} that I have cited in this and previous papers, is not widely used or very familiar to a large number of physicists.

Spinors like $\eta^A$ or $\pi^A$ belong to a simplectic complex two-dimensional vector space $S$. The complex conjugate vector space $S'$ has elements $\bar\eta^{A'}$. We also need to consider the two dual spaces $S^*,S^{*'}$ with elements $\xi_A,\bar\xi_{A'}$.
% 19.06.2018  Have inserted material from 2017 conference talk. I have modified the expression for %\eta^{A'} by putting bars on what I have assumed to be complex conjugates,  as I assume that \eta^{A'} is the complex conjugate % of \eta^{A}
\bea
\eta^A &  =  & \twospinor{\eta^0}{\eta^1}\,\, \in \,\, S, \qquad \,\,\, \bar \eta^{A'}\, =\,\, \twospinor{\bar \eta^{0'}}{\bar \eta^{1'}} \,\,\in \,\, S'  \nonumber    \\  
\xi_A  & =   &  \left( \,\xi_0 \quad\xi_1 \right) \,\,\in \,\,S^{\ast}, \qquad \bar \xi_{A'}\, = \,\, \left(\, \bar \xi_0 \quad \bar \xi_1 \right ) \,\, \in \,\, S^{\ast'}
\eea

Just as the familiar metric tensor $\eta^{\mu\nu}$ of Minkowski space, in $ S $ we also have a metric spinor
% 19.06.2018 Can save time/effort using the user defined commands 
%
%\be  \label{metric}
%\epsilon^{AB} = \epsilon^{A'B'} = \epsilon_{AB} = \epsilon_{A'B'} = \twomatrix{0}{1}{-1}{0}
%\ee
%
\begin{equation}\label{metric}
\epsilon_{AB}=\epsilon^{A'B'} = \epsilon_{AB} = \epsilon_{A'B'}\left(\begin{array}{cc} 0 & 1 \\ -1& 0 \end{array}\right)
\end{equation} \\
%19.06.18 Removed space between end of equation and text to remove unwanted indent at the beginning of the following line of text.  
relating any spinor $\eta^A$ with $\eta_A$ according to the rules:
$$
\eta^A=\epsilon^{AB}{\eta_B}
$$
and
$$
\xi_A=\epsilon_{BA}\xi^B,
$$
with similar rules for the complex conjugate quantities. It follows that the components of $\eta^{A}$ are related to the components of $\eta_{A}$ by  
$$
\eta^0 = \eta_1, \quad \eta^1 = - \eta_0 
$$
Accordingly, for any spinor 
$$ 
\eta^A\eta_A= \eta^0 \eta^1 - \eta^1 \eta^0 = 0
$$ 
Care is needed with the index ordering because $\epsilon_{AB}$ is skew:
% 19.06.2018  It would be helpful to expand the raising and lowering of indices as 
$$
\eta^A=\epsilon^{AB}\eta_B=-\epsilon^{BA}\eta_B.
$$
%
% 19.06.2018  Have replaced ``spinor'' with ``spinors'' 
Just as in  ordinary tensor calculus, spinors of higher rank, like those used in this work: $p^{AA'}$, $\phi_{AB}$ and $F_{ABA'B'}$, can be defined. In particular, there is an isomorphism between real four vectors in Minkowskian space-time and hermitian second rank spinors:
%
% 19.06.2018  Have removed second index on V as it is meant to denote a 4-vector (as far as I know) 
\begin{equation}
V^{\alpha} \longrightarrow V^{AA'}=\frac{1}{\sqrt 2}\left( \begin{array}{cc}
V^0+V^3& V^1+iV^2\\
V^1-iV^2 & V^0-V^3\\
\end{array}\right).
\end{equation}
The Lorentz norm equals one half the determinant of the above matrix. In particular, if the vector is null, like $\pi^A\bar\pi^{A'}$, it may be written as the outer product of a complex two-dimensional vector and its complex conjugate:
\begin{equation}
\pi^A\bar\pi^{A'}=\left( \begin{array}{cc}
\pi^0\bar\pi^{0'} & \pi^0\bar\pi^{1'}\\
\pi^1\bar\pi^{0'} & \pi^1\bar\pi^{1'}\\
\end{array}\right).
\end{equation}
\bigskip

{\noindent \bf Appendix 2: Classical Weyl-spinor lagrangian and quantum Weyl and Dirac lagrangians}
\bigskip

As we have seen the rather simple choice, in spinor language, of the lagrangian density leading to the spinor equation of motion is
$$
{\cal L}= \pi_A \dot\eta^A, \eqno(14b)
$$
in contrast, by comparison, with the familiar expression of the Dirac quantum lagrangian for a free particle

\begin{equation}
	{\cal L}= i\bar\psi \gamma^\mu \partial_\mu \psi - m \bar\psi \psi.
\end{equation}
%
%The lagrangian density given by (14b) might be though as the spinor analogue of \cite{landau}
%
%\begin{equation}
%L=-m\sqrt{1-v^2},
%\p%\end{equation}
%
%for a free relativistic but classical, as opposed to quantum, particle. Thus in the particle's rest frame becomes $L=m$ which is closely related to (14b). This may be seen by noting that $p^{AA'}p_{AA'}=m^2$ implies that either
%
%\begin{equation}
	%\pi^A\eta_A =m \Rightarrow \dot\pi^A\eta_A+\pi^A\dot\eta_A=0
%\end{equation}
%
%or
%
%\begin{equation}
%\pi_A\eta^A=-m \Rightarrow \dot\pi_A\eta^A+\pi_A\dot\eta^A=0.
%\end{equation}
% 23.08.2018   change ``spinor values quantities'' to ``spinor valued quantities'' and put ``$m$'' instead of ``m'' to denote the rest mass 
%We are free to choose (46). However the lagrangian itself cannot be equal to a constant, [5] including zero, as this would lead to Euler-Lagrange equations which are trivially satisfied by all spinor valued quantities $\eta^A$ and $\pi_A$. Since the rest mass, $m$, is not only an intrinsic property of the particle but also a Lorentz scalar, a lagrangian based on or derived from (45) or (46) would also be a Lorentz scalar and therefore a suitable choice. Hence we arrive at (14b).

%we see that although this equation gives the correct form for the momentum, it is not a tensor equation and depends on a frame dependent parameter (coordinate time $t$). A more convenient form is
%
%\begin{equation}
	%L=\frac{1}{2}mu^\alpha u_\alpha,
%\end{equation}
%
%with $u^\alpha=(\gamma,\gamma v^1,\gamma v^2, \gamma v^3)$.
%

One thing to note is that while the spinor lagrangian density has dimension energy per unit length, the dimension of the classical lagrangian is energy thus emphasizing  the deep difference between the standard approach and the spinorial one (note that the mass is also absent in the spinor definition of the four momentum). In any case it is apparent that classical lagrangians for a free particle only contain one term.

If we try to find the Weyl version of the Dirac Equation (DE) via the usual lagrangian approach we need to add a mass term. In [1] the Weyl 2-spinor version of the DE is found in a simple way starting from
$$
p^{AA'}=\frac{1}{\sqrt 2}[\pi^A\bar\pi^{A'}+\eta^A\bar\eta^{A'}]
$$
(see (22) and subsequent equations in [1]).

Following now the standard approach, in the DE case the Euler-Lagrange equations to obtain the DE from the lagrangian (43) are
\begin{equation}
	\partial_\mu \frac{\partial{\cal L}}{\partial\psi_{,\mu}}-\frac{\partial{\cal L}}{\partial\psi}=0,
\end{equation}
and a similar one for $\bar\psi$. As it is standard material found in many books, we shall not develop further the steps leading to the DE. However, in our Weyl two-spinor approach things are, if not substantially more involved, less familiar.
% 23.08.2018  Replace singular ``functional'' with plural ``functionals'' 
Let  us start with the free lagrangian (14b). Spinors $\pi^A$ and $\eta^A$ are now no longer defined along any classical path parameterized by proper time but instead should be regarded as functional or distributions in four dimensional spacetime. In consequence, the derivative with respect to proper time should be replaced by general derivatives in the context of two-spinor calculus:
\begin{equation}
	\dot\pi^A \rightarrow \nabla^{AA'}\bar\pi_{A'}
\end{equation}
\begin{equation}
	\dot{\bar \eta}_{A'} \rightarrow \nabla_{AA'}\eta^A.
\end{equation}
% 23.08.2018 Replace ``adequate'' with ``appropriate'' 
The appropriate Euler-Lagrange equations in two-spinor calculus (generalization of (15)) is
\begin{equation}
	\nabla^{AA'}\frac{\partial \cal L}{\partial(\nabla^{AA'}\bar\pi_{A'})}-\frac{\partial\cal L}{\partial \bar\pi_{A'}}=0
\end{equation}
and
\begin{equation}
	\nabla_{AA'}\frac{\partial \cal L}{\partial(\nabla_{AA'}\eta^A)}-\frac{\partial\cal L}{\partial \eta^A}=0.
\end{equation}
The two coupled lagrangians are defined as
\begin{equation}
     {\cal L}=\pi_A \nabla^{AA'} \bar\pi_{A'} +\frac{m}{\sqrt 2} \eta^{A'}\bar\pi_{A'}
\end{equation}

\begin{equation}
	{\cal L}= \bar\eta^{A'} \nabla_{AA'}\eta^A -\frac{m}{\sqrt 2} \eta^A \pi_A.
\end{equation}
Taking the first
$$
\frac{\partial\cal L}{\partial(\nabla^{AA'}\bar\pi_{A'})}=\pi_A 
$$
$$
\frac{\cal L}{\partial\pi_A'}=\frac{m}{\sqrt 2}\eta^{A'},
$$
and from (50):
$$
\nabla^{AA'}\pi_A=\frac{m}{\sqrt 2}\eta^{A'}.
$$
Performing the same calculations with the second we arrive at the 2-spinor version of the DE in Weyl representation as they appear in \cite{penrose}:
\begin{equation}
	\nabla^{AA'}\pi_A=\frac{m}{\sqrt 2}\bar\eta^{A'}
\end{equation}
\begin{equation}
	\nabla_{AA'}\bar\eta^{A'}=-\frac{m}{\sqrt 2}\pi_A.
\end{equation}
\bigskip
%
%A theoretical article should be restrained to the essential content and devoid of any personal or historical digression so that I think that the following paragraphs should not appear (I keep them in the source manuscript for own purposes only)
%
\bigskip


\begin{thebibliography}{99}

\bibitem{bui} Buitrago J. Results in Physics 6 (2016) 346-351
%
\bibitem{hestenes} Hestenes D. Found. Physics., Vol. 20, No. 10, (1990) 1213

\bibitem{bette} A. Bette, J. Math. Phys. $\mathbf{34}$, 4617 ($1993$)
%
\bibitem{sch} Schrodinger E. Sitzungber. Preuss. Akad. Wiss. Phys.-Math. Kl. 24, 418 (1930)
%
\bibitem{buitrago} J. Buitrago and S. Hajjawi \textit{Spinor extended Lorentz-force-like equation as 
a Consequence of a Spinorial Structure of Space-Time}, J.Math.Phys., $\b{48}$ $022902$ ($2007$).
%
\bibitem{landau} L.D. Landau and E.M. Lifshitz, {\it The classical Theory of Fields}, Pergamon Press, 1971.
%
\bibitem{penrose} R. Penrose and W. Rindler \textit{Spinors and Space-Time}, 
Cambridge Monographs in Mathematical Physics, Vol. $1$, Cambridge Universtiy Press, Cambridge, 
England ($1984$/$1986$).
%
\bibitem{bjork} J.D. Bjorken and S.D. Drell {Relativistic Quantum Mechanics} McGraw Hill (1964)
%
\bibitem{bui0} J. Buitrago Eur J Phys  16, 113 (1995)
%
\bibitem{physreva} W.E. Baylis and Y. Yao Phys Rev A 60, 2 (1999)
%
%\bibitem{griff} D. Griffiths \textit{Introduction to Elementary Particles} John Wiley and Sons, Inc ($1987$)
%
\end{thebibliography}
\end{document}